\documentclass[aps,pra,twocolumn,showpacs,superscriptaddress,10p,longbibliography]{revtex4-1}
\usepackage{graphicx}
\usepackage{epsfig}
\usepackage{dcolumn}
\usepackage{amsmath}

\usepackage[sort&compress]{natbib}

\usepackage{bm}
\usepackage{bbm}
\usepackage{ulem}
\usepackage[colorlinks, citecolor=red]{hyperref}
\usepackage{mathtools}
\usepackage{color}
\usepackage{soul}
\usepackage{centernot}
\usepackage[up]{subfigure}
\usepackage{dsfont}	
\usepackage{relsize}
\usepackage{epstopdf}
\usepackage{mathrsfs}
\usepackage{braket}
\begin{document}
\title{Quench dynamics in disordered two-dimensional Gross-Pitaevskii Lattices}

   \author{Arindam Mallick}
   \email{marindam@ibs.re.kr}
  \affiliation{Center for Theoretical Physics of Complex Systems, Institute for Basic Science (IBS), Daejeon 34126, Republic of Korea}
  
  \author{Thudiyangal Mithun}
  \email{mthudiyangal@umass.edu}
  \affiliation{Center for Theoretical Physics of Complex Systems, Institute for Basic Science (IBS), Daejeon 34126, Republic of Korea}
  \affiliation{Department of Mathematics and Statistics, University of Massachusetts, Amherst, MA 01003-4515, USA}
  
    \author{Sergej Flach}
    \email{sflach@ibs.re.kr}
  \affiliation{Center for Theoretical Physics of Complex Systems, Institute for Basic Science (IBS), Daejeon 34126, Republic of Korea}
\begin{abstract}
We numerically investigate the quench expansion dynamics 
of an initially confined state  in a two-dimensional Gross-Pitaevskii lattice in the presence of external disorder. 
The expansion dynamics is conveniently described in the control parameter space of the energy and norm densities.
The expansion can slow down substantially if the expected final state is a non-ergodic non-Gibbs one, regardless of the disorder strength. Likewise stronger disorder delays expansion.
We compare our results with recent studies for quantum many body quench experiments.
\end{abstract}

\maketitle


\section{Introduction}
Quench dynamics is a common way to explore the cooling process of nonequilibrium states in Hamiltonian systems. 
It implicitly assumes the ability of the system to thermalize and equilibrate.
The quench dynamics is particularly important when investigating
localization-delocalization phenomena and the related presence or absence of thermalization \cite{Polkovnikov_2011}. 
Quench dynamics is therefore also widely used 
for measuring the different time scales involved in 
a thermalization process.

Recent experiments with interacting ultracold bosonic atomic gases loaded into two-dimensional disordered optical potentials
used the quench dynamics to  explore the signatures of the many-body localization-delocalization
transition \cite{Choi1547}. The atomic gas was confined and prepared in a thermal state and then allowed to expand into a previously empty part of the random potential. Localization-delocalization transitions were observed upon varying the disorder strength and the atom-atom interaction
strength. Subsequent computational studies with quantum many-body platforms using Gutzwiller mean field methods \cite{Yan_2017} and
tensor network methods \cite{urbanek2018}
pointed to a number of open questions such as the impact of
the system size and measurement times.

The dynamics of ultracold bosonic atoms in a deep optical lattice can be modeled with a Bose-Hubbard Hamiltonian (BH). 
For sufficiently large occupation numbers its classical counterpart---the discrete Gross-Pitaevskii (DGP) Hamiltonian---serves as a reasonable approximation \cite{Dutta_2015}. 
The experimental studies of Choi {\it{et al.}}~were performed deep in the quantum regime with at most double occupancy per lattice site (see supplement of Ref. \cite{Choi1547}). Despite that discrepancy, the merit in the DGP approach is that large systems can be evolved up to large times using
standard computational approaches and average computational resources. 
The DGP Hamiltonian is  
also known as the discrete nonlinear Schr\"{o}dinger (DNLS) Hamiltonian \cite{kevrekidis2009discrete} and serves as a platform to study various
properties of nonlinear wave dynamics.

Many body localized phases are expected to be non-ergodic and non-thermalizing \cite{RevModPhys.91.021001}, at variance to their delocalized (metallic) counterparts.
Many body localized phases are as well expected to be unique for quantum many body dynamics, at variance to classical wave dynamics.
Therefore the DGP model can be expected not to possess a many-body localization-delocalization
transition. However, the classical DGP model, as well as its quantum BH counterpart, exhibit a non-Gibbs phase, which is 
characterized by at least partial nonergodic properties and absence of full thermalization \cite{Mithun_2018,PhysRevA.99.023603}. 
An intriguing question is therefore whether
these non-Gibbs phases have an impact on the outcome of the quench dynamics. 

The article is organized as the following. In section \ref{model} we introduce the DGP model and its statistical description. 
In section \ref{quench_dynamics} we present our results on the quench dynamics of the DGP. In the section \ref{comparison} 
we compare our numerical results with the experimental results reported in \cite{Choi1547}. The section \ref{concl} concludes and discusses
the results.

\section{The Model}\label{model}

We consider the following two-dimensional DGP Hamiltonian in dimensionless unit
\begin{align}\label{refhamil}
 \mathcal{H}  = & \sum_{n} \sum_{m}  \frac{U}{2} |\psi_{m,n}(t)|^4 +  V_{m,n} |\psi_{m,n}(t)|^2 \nonumber\\
  & -  J  \Big[ \psi^*_{m,n}(t) \psi_{m+1,n}(t) + \psi_{m,n}(t) \psi^*_{m+1,n}(t) \nonumber \\
  & + \psi^*_{m,n}(t) \psi_{m,n+1}(t) 
 + \psi_{m,n}(t) \psi^*_{m,n+1}(t) \Big],
\end{align}
 where $J$ is the hopping strength, ($\psi_{m,n}(t)$, $\psi^{\ast}_{m,n}(t)$) represent the conjugated variables and the indices ($m$, $n$) represent
 the lattice sites in a square lattice. 
 Here
$U$ is the nonlinearity parameter and $V_{m,n}$ represents the uncorrelated onsite disorder potential of the form 
\begin{align}
 V_{m,n} & =  \epsilon_{m,n} ~&\text{for}~1 \leq m \leq L, 1 \leq n \leq L; \nonumber\\
         & =  \infty  ~&\text{otherwise}.
\end{align}
The uncorrelated onsite energies $\epsilon_{m,n}$ are
taken from a uniform distribution with the range $\in$ $\left[ -\frac{W}{2}, \frac{W}{2}\right]$. 
This potential enforces fixed boundary conditions
$\psi_{m,n} = 0$ outside the boundary ($m = 1, L$; $n = 1, L$) at all times $t$. 

The Hamiltonian, Eq.~\eqref{refhamil} gives the following equations of motion
\begin{align}
 i \frac{\partial}{\partial t}  \psi_{m,n}(t) 
  = U |\psi_{m,n}(t)|^2 \psi_{m,n}(t)  + V_{m,n} \psi_{m,n}(t) \hspace{1cm}\nonumber\\
 -J \Big[ \psi_{m+1, n}(t) + \psi_{m-1,n}(t)
 + \psi_{m, n+1}(t) + \psi_{m, n-1}(t) \Big].
 \label{Eq. dgpe}
\end{align}
Eq.~\eqref{Eq. dgpe} possesses two conserved quantities, the total norm $ \mathcal{N}$ = $\sum_{m,n} |\psi_{m,n}|^2$ and the total energy $\mathcal{H}$.
Corresponding to the two conserved quantities, we define the norm density $a = \frac{\mathcal{N}}{L^2} $ and the energy density $h = \frac{\mathcal{H}}{L^2}$.  In the absence of nonlinearity $U=0$ and disorder the solutions are plane waves $\exp [i(k_m m+k_n n-\omega_k t)]$ with
$\omega_k = -2J(\cos k_m + \cos k_n)$. It follows that the linear system (even with disorder) has a spectrum of eigenfrequencies (or eigenenergies)
whose width  amounts to $\Delta \omega = 8J + W$.

If the microcanonical dynamics generated by (\ref{Eq. dgpe}) is ergodic, then infinite time averages of observables are equal to their phase space averages,
and the statistical properties of the system can be described using the Gibbs grand-canonical partition function
\begin{align}
 Z = \int e^{-\beta ( \mathcal{H}+\mu \mathcal{N})} \prod_{m = 1}^{L} \prod_{n = 1}^{L} d \psi_{m,n} d \psi_{m,n}^{\ast}.
 \label{eq:partition}
\end{align}  
Here $\beta$ is the inverse temperature and $\mu$ is chemical potential. 
It follows that the density pair $\{a,h\}$ can be mapped onto a pair of Gibbs parameters $\{\mu,\beta\}$ and vice versa.
In the following we will use scaled densities $x=Ua$ and $y=Uh$.
Since the seminal publications \cite{Rasmussen_2000, Johansson_2004} it is known, that the one-dimensional ordered discrete nonlinear
Schr\"odinger lattice has a groundstate line $y_0(x)$ on which the temperature vanishes $\beta^{-1}=0$. At the same time there is a second
line $y_{\infty}(x) = x^2 > y_0(x)$ on which the temperature diverges $\beta=0$. All microcanonical states $y(x) > y_{\infty}(x)$ can not
be described by a Gibbs distribution with a positive temperature, and negative temperature assumptions lead to a divergence of the
partition function (technically this happens only on infinite systems; we will assume here that our considered system sizes are large enough
for this statement to apply). Recently these results were generalized to Gross-Pitaevskii lattices with any lattice dimension and disorder,
and even to corresponding quantum many-body interacting Bose-Hubbard lattices \cite{PhysRevA.99.023603}.
While the zero-temperature line $y_0(x)$ renormalizes in the presence of a disorder potential,
the infinite temperature line $y_{\infty}(x) = x^2$ is invariant under the addition of disorder.

 \begin{figure}[!htbp]
 \subfigure{\includegraphics[width = 8.5cm]{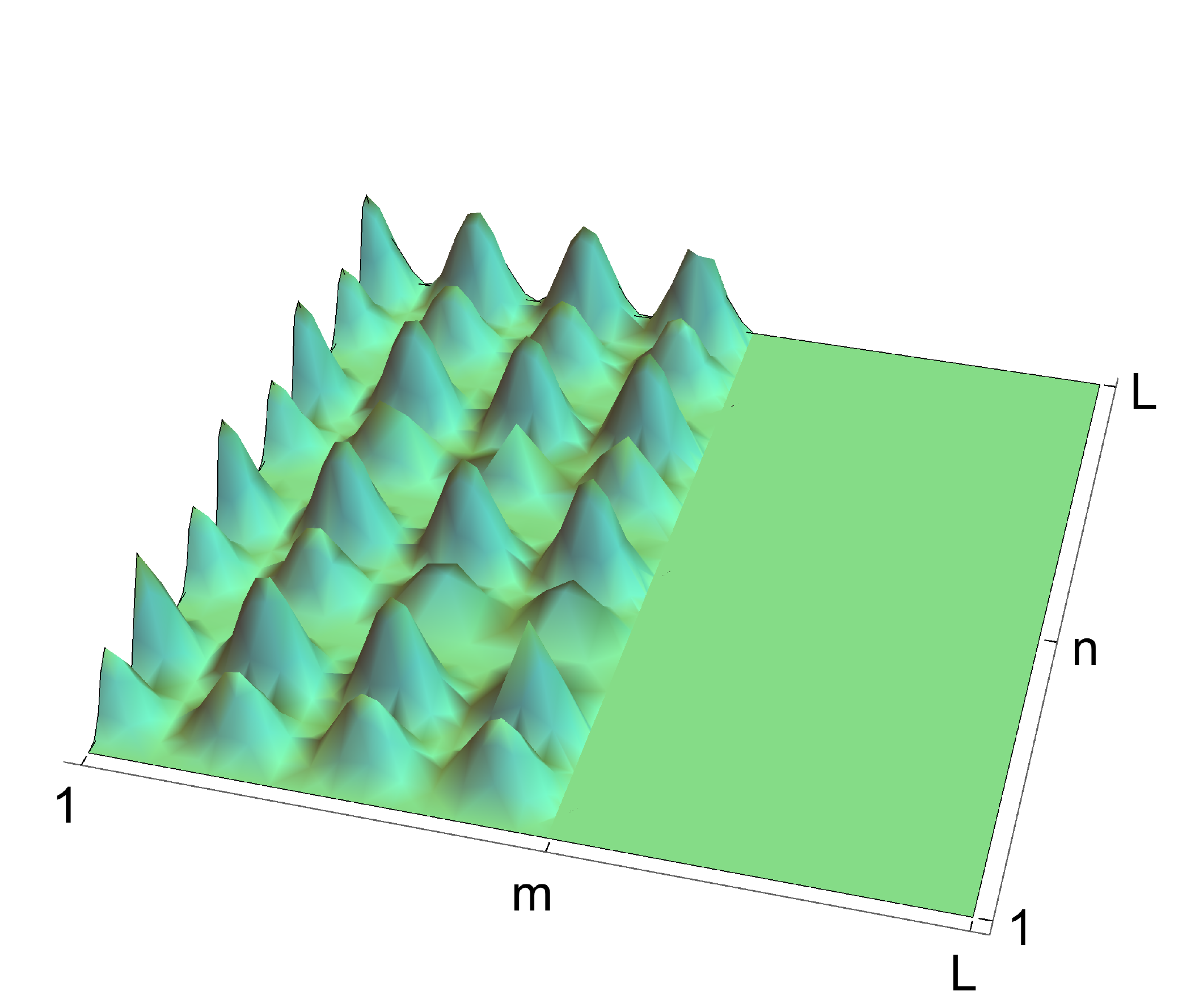}}
  \caption{Schematic distribution of the initial wave function norm  $|\psi_{m,n}|^2$ profile 
  on a square lattice of size $ L \times L$ with fixed boundary conditions. Inside the square lattice 
  the initial wavefunction is strictly zero for $m > L/2$.
}
  \label{initial_profile}
  \end{figure}

  We use a symplectic scheme \cite{Yoshida:1990aa,McLachlan:1995aa,laskar_2001} to numerically integrate Eq.~\eqref{Eq. dgpe}. 
  The details of the symplectic integration method $\mathcal{SBAB}_2$ can be found in
  Refs.~\cite{Skokos:2009aa,Skokos:2014aa,Carlo_2019}. We consider time steps $\Delta t=0.005$ to keep the relative error in energy $\Delta H =(H(t)-H(0))/H(0)$ 
  and norm $\Delta \mathcal{N} =(N(t)-N(0))/N(0)$ smaller than $10^{-3}$.

\section{Quench dynamics}\label{quench_dynamics} 
We consider a square lattice of size $L \times L$ with $L=16$. We set the total norm $\mathcal{N}= 125$ in loose analogy to the experiments \cite{Choi1547} which 
trapped 125 atoms. Thus roughly one unit of norm in our numerical experiments corresponds to one atom.
We prepare an initial state of plane waves $\psi_{m,n}(t = 0)$ = $ \sqrt{a_0} e^{i \phi_{m,n}(t = 0)}$ if $m$ $\in$ $\left\{1,2,3, \ldots, \big\lfloor\frac{L}{2} \big\rfloor \right\}$ occupying
one (left) half of the system $\mathcal{L}$, i.e. $\psi_{m,n} = 0$ for $m$ $\in$ $\left\{\big\lfloor\frac{L}{2} \big\rfloor + 1, \big\lfloor\frac{L}{2} \big\rfloor + 2, \ldots, L \right\}$ in the right half of the system $\mathcal{R}$. 
Fig.~\ref{initial_profile} shows the schematic representation of the initial state.
The initial norm density in the excited half  $\mathcal{L}$ of the system is $a_0=\frac{125}{L^2/2} \approx 0.98$ (before the quench).
If the excitation spreads over the entire system,
the expected final norm density in the entire system (after the quench) becomes $a = \frac{125}{L^2} \approx 0.49$. 

We follow the evolution of the local norm density $|\psi_{m,n}|^2$. 
In addition to the real space imaging of $|\psi_{m,n}|^2$ at the final time, we measure the time evolution of the left-right norm imbalance ratio:
\begin{align}
I(t) = \frac{\sum_{(m,n) \in \mathcal{L} } |\psi_{m,n}(t)|^2 
- \sum_{(m,n) \in \mathcal{R}}  |\psi_{m,n}(t)|^2}
{\sum_{(m,n) \in \mathcal{L}} |\psi_{m,n}(t)|^2 + \sum_{(m,n) \in \mathcal{R} }  |\psi_{m,n}(t)|^2}.
\end{align}
The imbalance is bounded by $|I| \leq 1$. At $t=0$ it follows $I(0) = 1$. 
Further, at equilibrium  $\sum_{(m,n) \in \mathcal{L} } |\psi_{m,n}|^2 = \sum_{(m,n) \in \mathcal{R}}  |\psi_{m,n}|^2 $. 
Hence after some equilibration time $t_\text{eq}$ the 
norm imbalance practically vanishes $I(t \geq t_\text{eq})$ $\approx$ 0.

In the absence of nonlinearity, $U = 0$, the Eq.~(\ref{Eq. dgpe}) is integrable and analytically solvable. 
For the linear ordered case $U = 0$, $W = 0$ a set of plane waves appear as the eigenfunctions. 
In this case it follows that the imbalance ratio will show large amplitude oscillations  with time, without any tendency to thermalize
and diminishing of the oscillation amplitudes. 
In presence of disorder,
$W \neq 0$ the system shows Anderson localization \cite{Anderson_1958}. The initial state will not propagate into the entire system, and the 
imbalance $I$ will saturate at some nonzero value depending on $W$.
The presence of nonlinearity destroys integrability. This will usually lead to a restoring of ergodicity, and thermalization. 
Consequently the imbalance is expected to saturate at value zero. 
At variance to classical field equations, quantum many body interacting systems can show many-body localization phases which withstand the above scenario 
\cite{RevModPhys.91.021001}, so that the imbalance is expected to saturate at a nonzero
value. This precise prediction was tested in the experiments on cold atoms \cite{Choi1547}. However, the DGP system while being classical also possesses nonergodic phases
as discussed above. 
In order to study the impact of the nonergodic DGP phase on the quench dynamics, we will study the quench dynamics in the regime of weak nonlinear interactions
$x\ll 1$, strong nonlinear interactions $x \gg 1$, and for strong nonlinear interactions tuned close
to the experimental parameters in Ref.~\cite{Choi1547}.

\subsection{Quench dynamics in the Gibbs regime} \label{general study}

\begin{figure}[!htbp]
\subfigure{\includegraphics[width = 8.5cm]{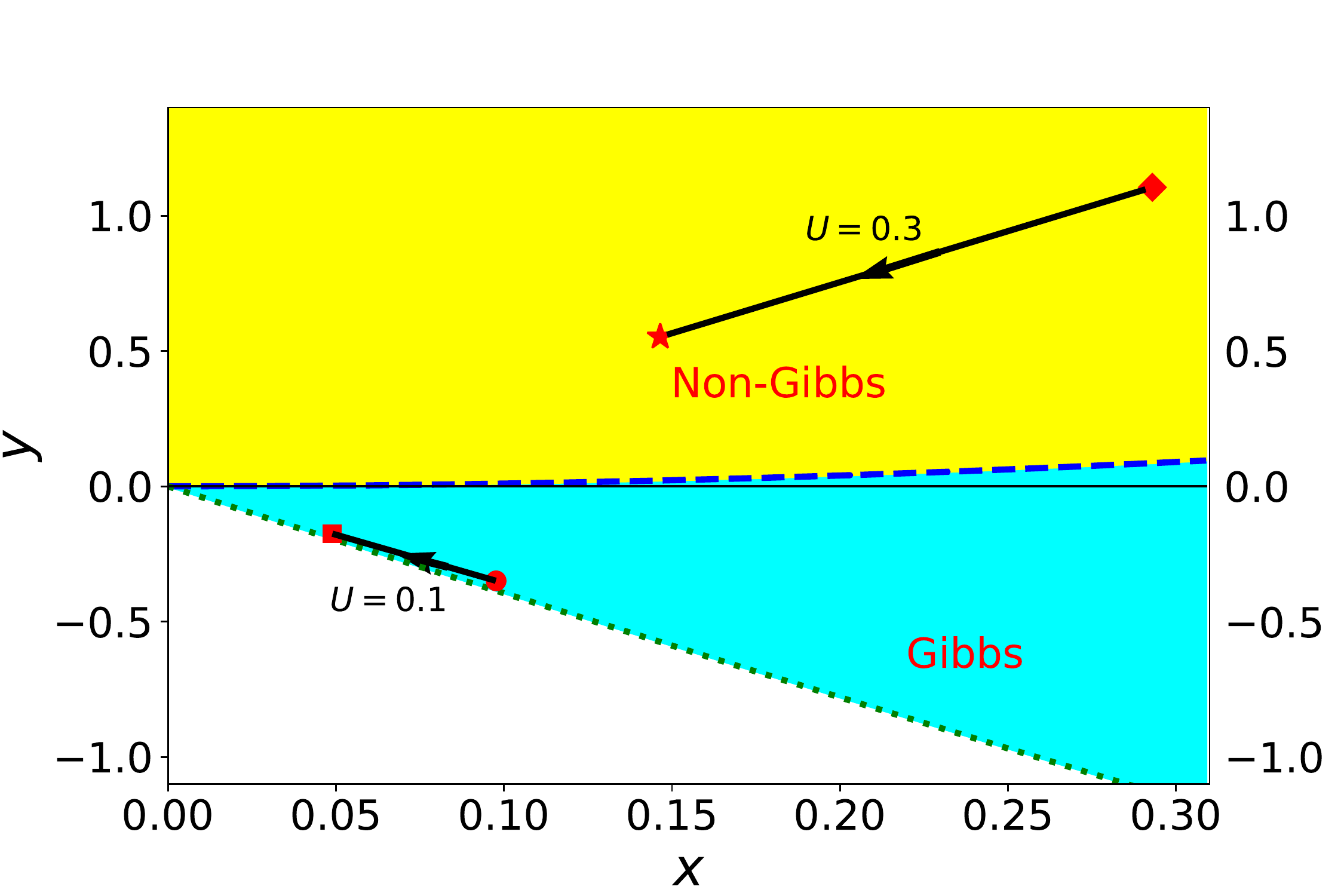}}
\caption{Small Nonlinearity $x \ll 1$: Phase diagram for the ordered case in the density parameter space $(x,y)$. 
The blue dashed curve is the transition line $y=x^2$ between the Gibbs (cyan) and non-Gibbs (yellow) regimes ($\beta = 0$). The green dotted line
is the ground state line for the ordered system  $y=-4x+x^2/2$ ($\beta = \infty$).
Each pair of symbols connected by lines with arrows denotes an initial state (larger norm density $x$) and the expected final state after the quench (smaller norm density $x$).
The corresponding values of $U=0.1$ and $U=0.3$ are denoted right to the pair lines.
}
\label{small_nonlinearity}
\end{figure}
 \begin{figure}[!htbp]
\subfigure{\includegraphics[width = 8.5cm]{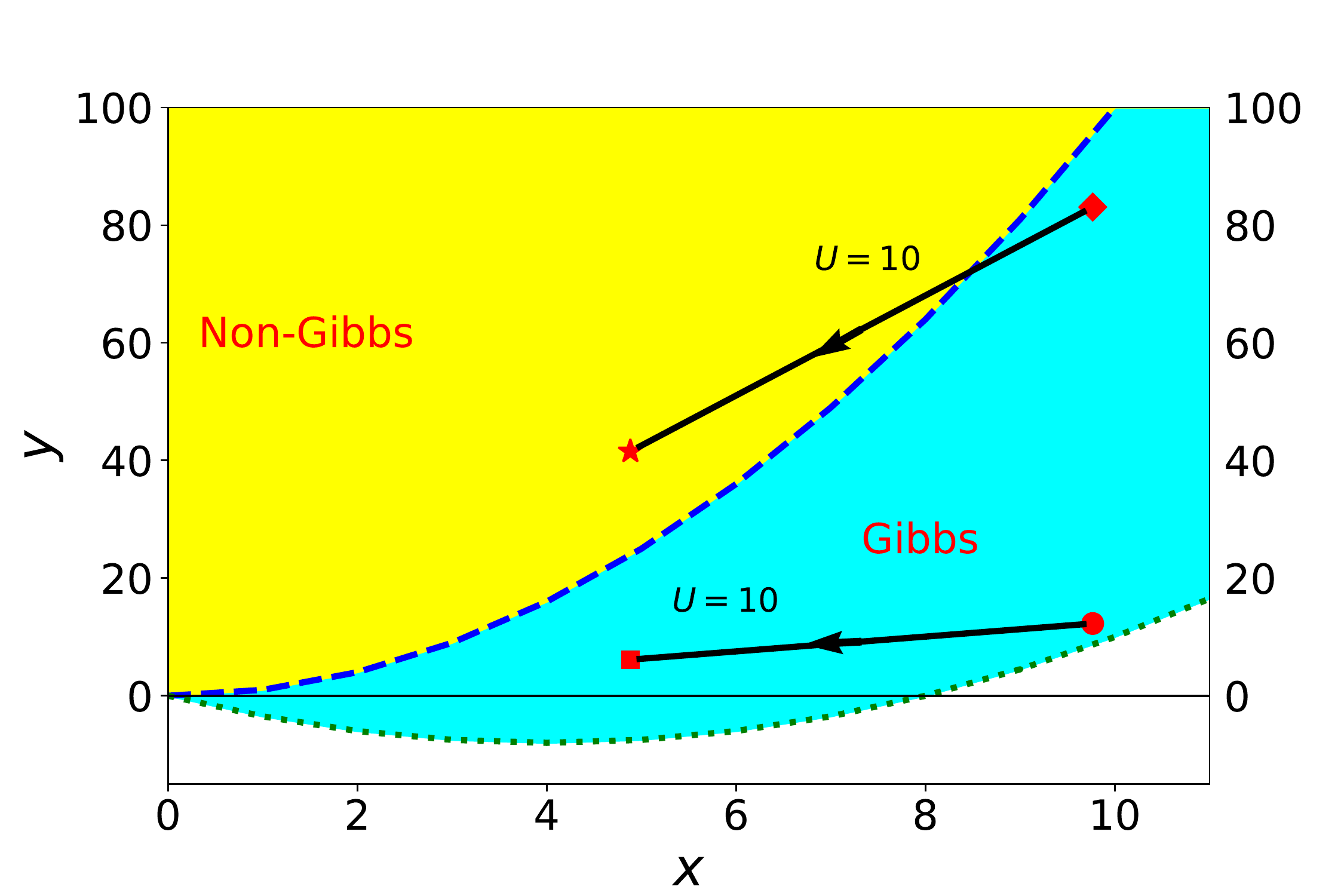}}
\caption{Large nonlinearity $x \gg 1$. Notations are as in Fig.~\ref{small_nonlinearity}.
}
\label{large_nonlinearity}
\end{figure}

We first consider quenches which start and end in the Gibbs regime. 
We use $\phi_{m,n}(t = 0) = 0$. This choice starts the dynamics close to the ordered system ground state line $y=-4x+x^2/2$  and 
keeps the system in the Gibbs regime after the quench, irrespective of the value of $U$.
   \begin{figure}[!htbp]
\subfigure[]{\includegraphics[width = 8.5cm]{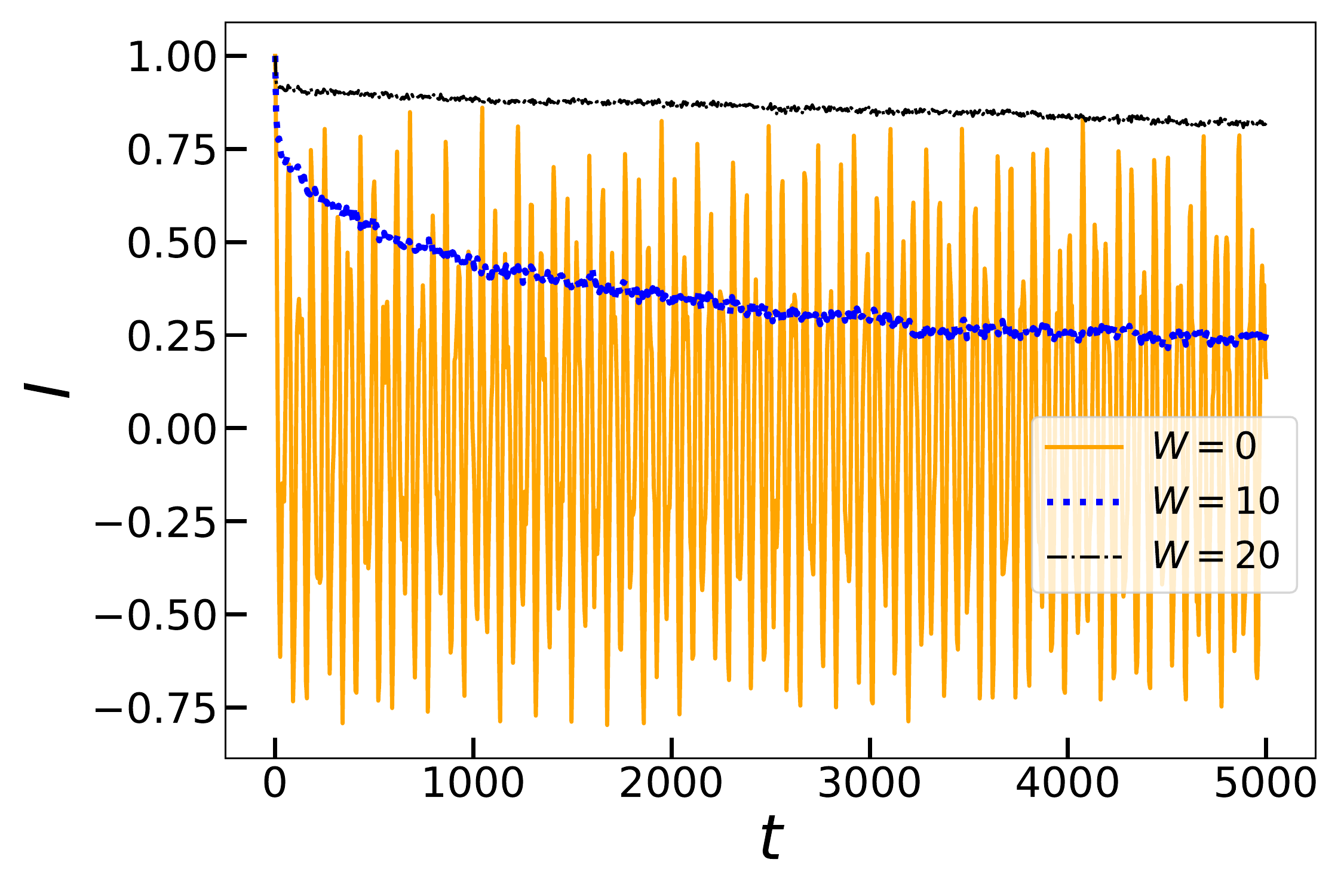}}\\
   \subfigure{\includegraphics[width = 9 cm]{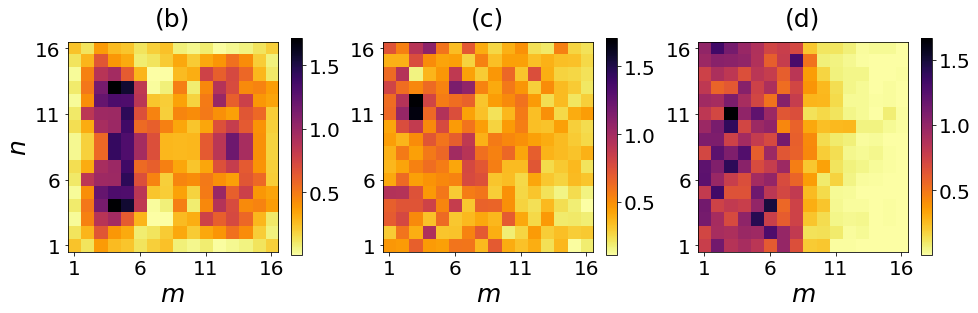}}
     \caption{
Quench dynamics for the $U=0.1$ path in Fig.~\ref{small_nonlinearity}.
(a) Imbalance $I(t)$ for $W=0,10,20$. (b)-(d) Norm density distribution at final time $t=5000$.
(b) $W=0$, (c) $W=10$, (d) $W=20$. 
All data for $W \neq 0$ are averaged over 20 disorder realizations.}
         \label{nu_0_1_gibbs}
    \end{figure}
For weak nonlinearity $U=0.1$ the quench line is shown in 
Fig.~\ref{small_nonlinearity} to connect the red circle and square.  
The evolution outcome is shown in
Fig.~\ref{nu_0_1_gibbs} for three different values of disorder strength $W = 0, 10, 20$. 
As expected, for $W=0$ the imbalance $I(t)$ shows non-decaying large amplitude oscillations around zero.  
It indicates absence of thermalization of the system up to the final evolution time, as also seen from the final time density 
plot snapshot in Fig.~\ref{nu_0_1_gibbs}{\color{red}(b)}. As $W$ increases,
Anderson localization prevails on the time scales of the runs. The imbalance decay is slowing down and nearly saturates during the later time of evolution for $W=10,20$.
The snapshots of the final time density plots 
in Figs. \ref{nu_0_1_gibbs}{\color{red}(c)} and \ref{nu_0_1_gibbs}{\color{red}(d)} confirm the above findings.
 \begin{figure}[!htbp]
\subfigure[]{\includegraphics[width = 8.1cm]{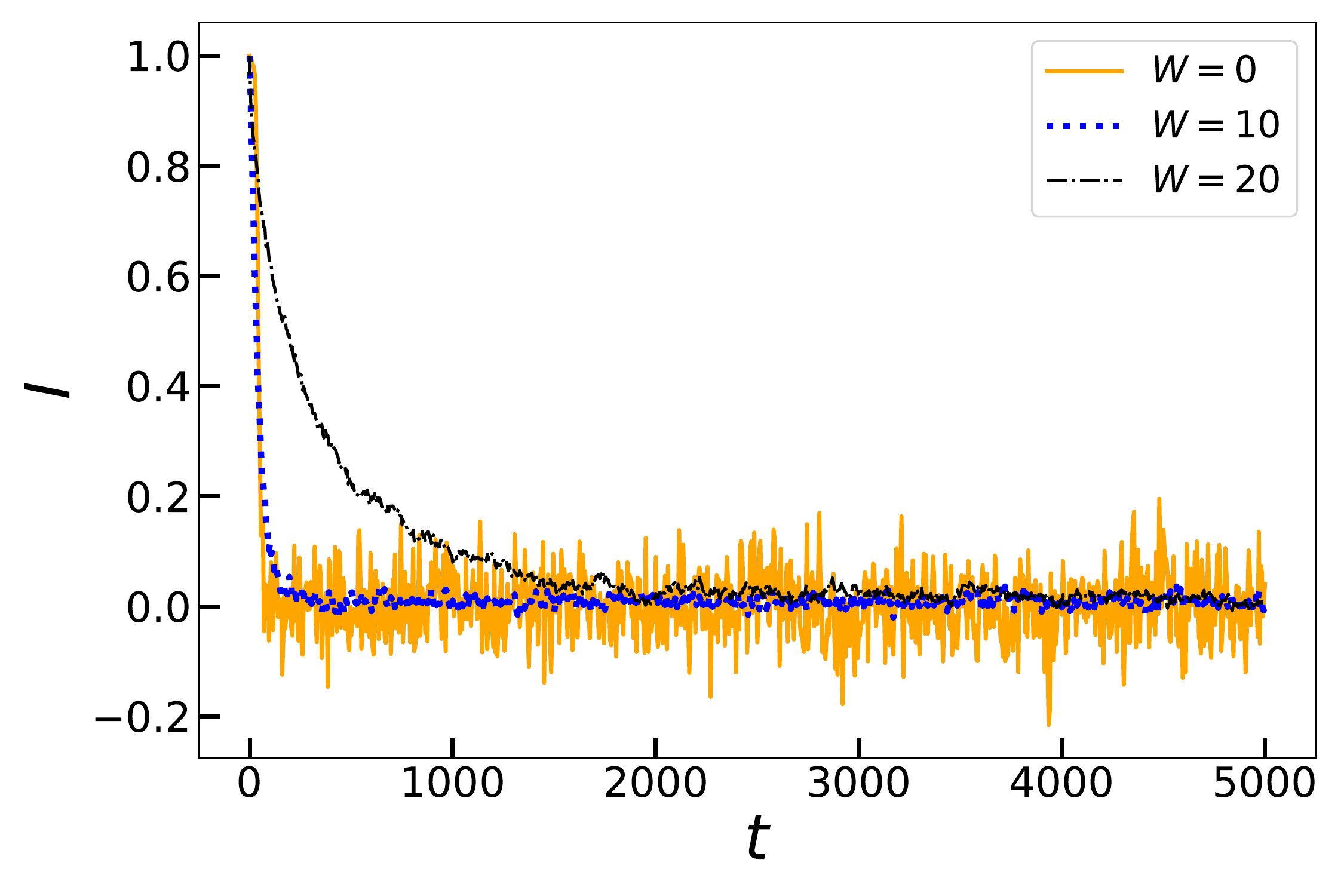}}\\
  \subfigure{\includegraphics[width = 9 cm]{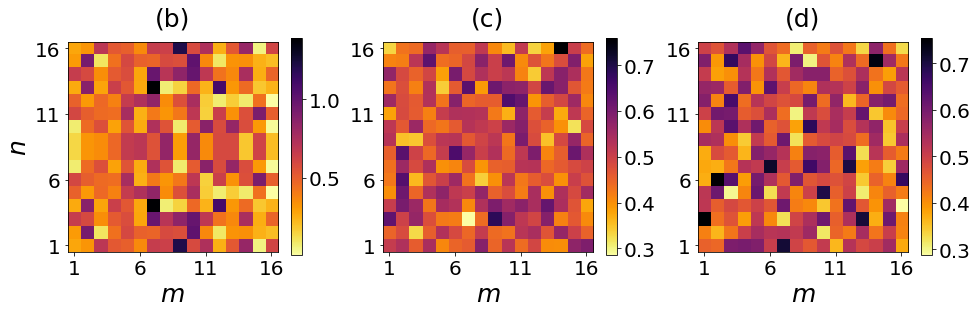}}
    \caption{
Quench dynamics for the $U=10$ path in Fig.~\ref{large_nonlinearity}.
(a) Imbalance $I(t)$ for $W=0,10,20$. (b)-(d) Norm density distribution at final time $t=5000$.
(b) $W=0$, (c) $W=10$, (d) $W=20$. 
All data for $W \neq 0$ are averaged over 20 disorder realizations.
}\label{nu_10_gibbs}
   \end{figure}

For strong nonlinearity $U=10$ the quench line is shown in 
Fig.~\ref{large_nonlinearity} to connect the red filled diamond and star. We observe thermalization and a decay of the imbalance to zero for all values of disorder $W=0,10,20$ in
Fig.~\ref{nu_10_gibbs}.
The thermalized density clouds at the final simulation times are shown in 
  Figs.~\ref{nu_10_gibbs}{\color{red}(b, c, d)}.


\subsection{Quench Dynamics in the Non-Gibbs regime}

\begin{figure}[!htbp]
\subfigure[]{\includegraphics[width = 8.2cm]{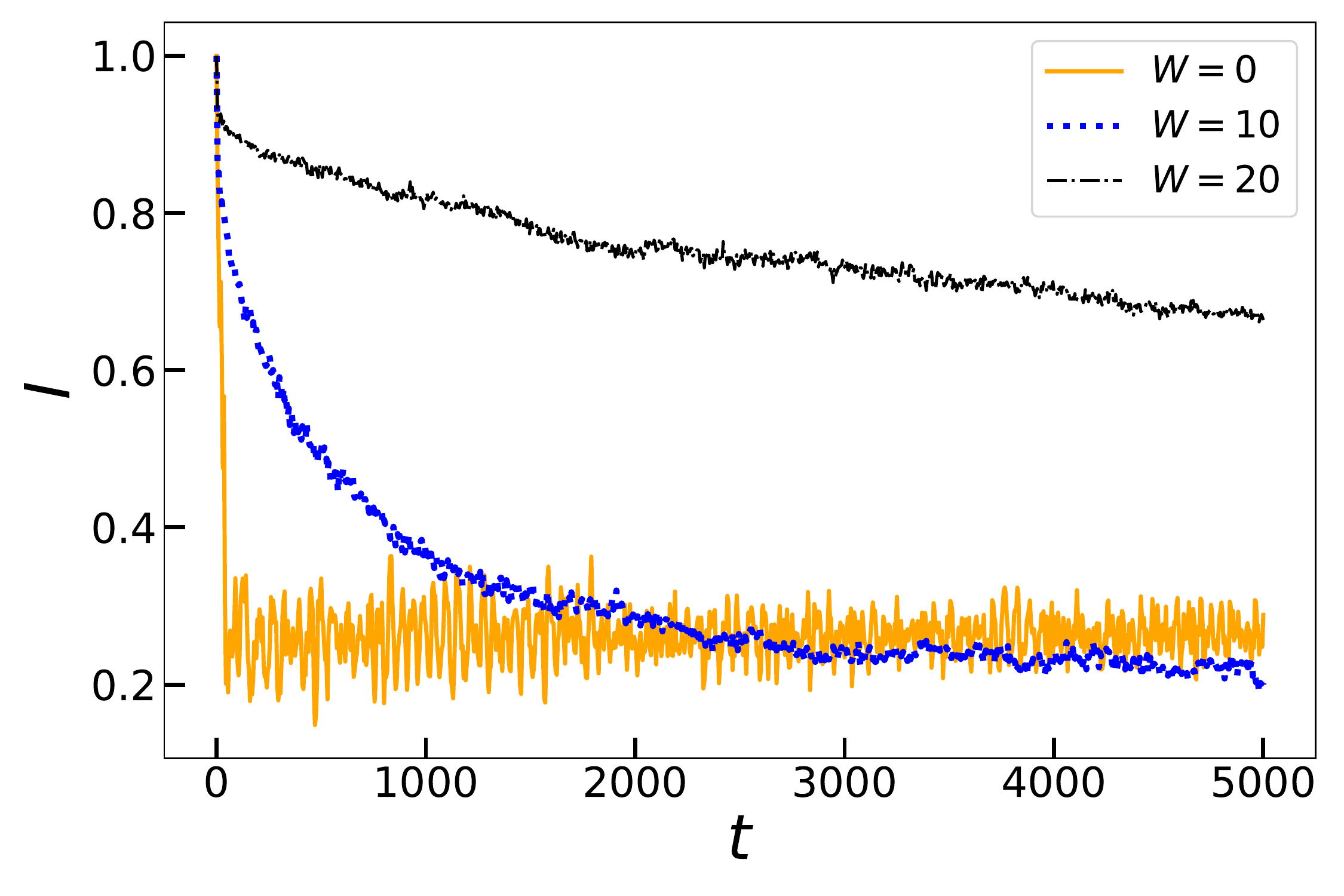}}\\
  \subfigure{\includegraphics[width =  9cm]{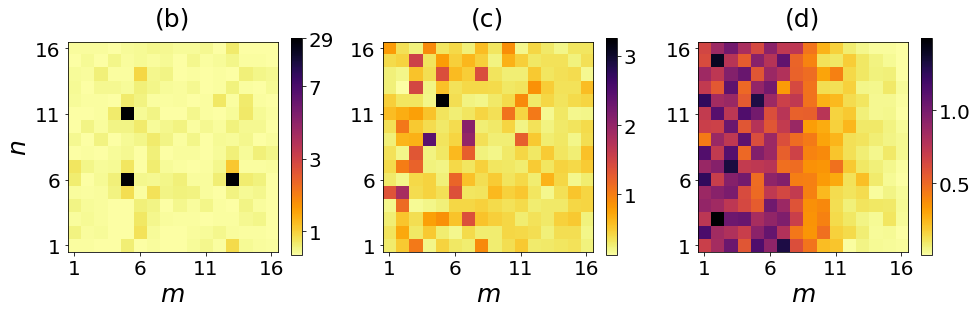}}
    \caption{
Quench dynamics for the $U=0.3$ path in Fig.~\ref{small_nonlinearity}.
(a) Imbalance $I(t)$ for $W=0,10,20$. (b)-(d) Norm density distribution at final time $t=5000$.
(b) $W=0$, (c) $W=10$, (d) $W=20$. 
All data for $W \neq 0$ are averaged over 20 disorder realizations.
}\label{nu_0_3_phase_pi}
   \end{figure}

We consider quenches which either start in the non-Gibbs regime and therefore stay in it, or which start in the Gibbs regime, but transit into the non-Gibbs one.
We initialize our system wavefunction with phases $\phi_{m,n}(t = 0) = \pi(m+n)$ 
so that the phase difference between any two nearest lattice neighbors is $\pi$.
For weak nonlinearity $U=0.3$ the quench path connects the triangle and the cross in 
Fig.~\ref{small_nonlinearity}.
The evolution outcome is shown in Fig.~\ref{nu_0_3_phase_pi}.
For $W=0$ we observe the formation of three persistent long-lived strongly localized large amplitude excitations Fig.~\ref{nu_0_3_phase_pi}{\color{red}(b)}. 
Each of them
confines a norm of about 30, which leaves a norm of about 35 to the background 
(barely visible).
Since two of the peaks are located in the left part and one in the right, the 
imbalance should take a value of about $30/125=0.24$ assuming that the background thermalizes. The dependence $I(t)$ in Fig.~\ref{nu_0_3_phase_pi}{\color{red}(a)} nicely confirms
these findings. Note that previous studies have observed and discussed the condensation of excess norm into strongly localized excitations such that the background will evolve
at an infinite temperature \cite{Rasmussen_2000,Johansson_2004, Rumpf_2004}. Increasing the strength of disorder to $W=10$ we still observe remnants of this non-Gibbs dynamics, while
even stronger disorder $W=20$ reinforces Anderson localization features.

 \begin{figure}[!htbp]
  ~~~~~~\subfigure[]{\includegraphics[width = 8.2cm]{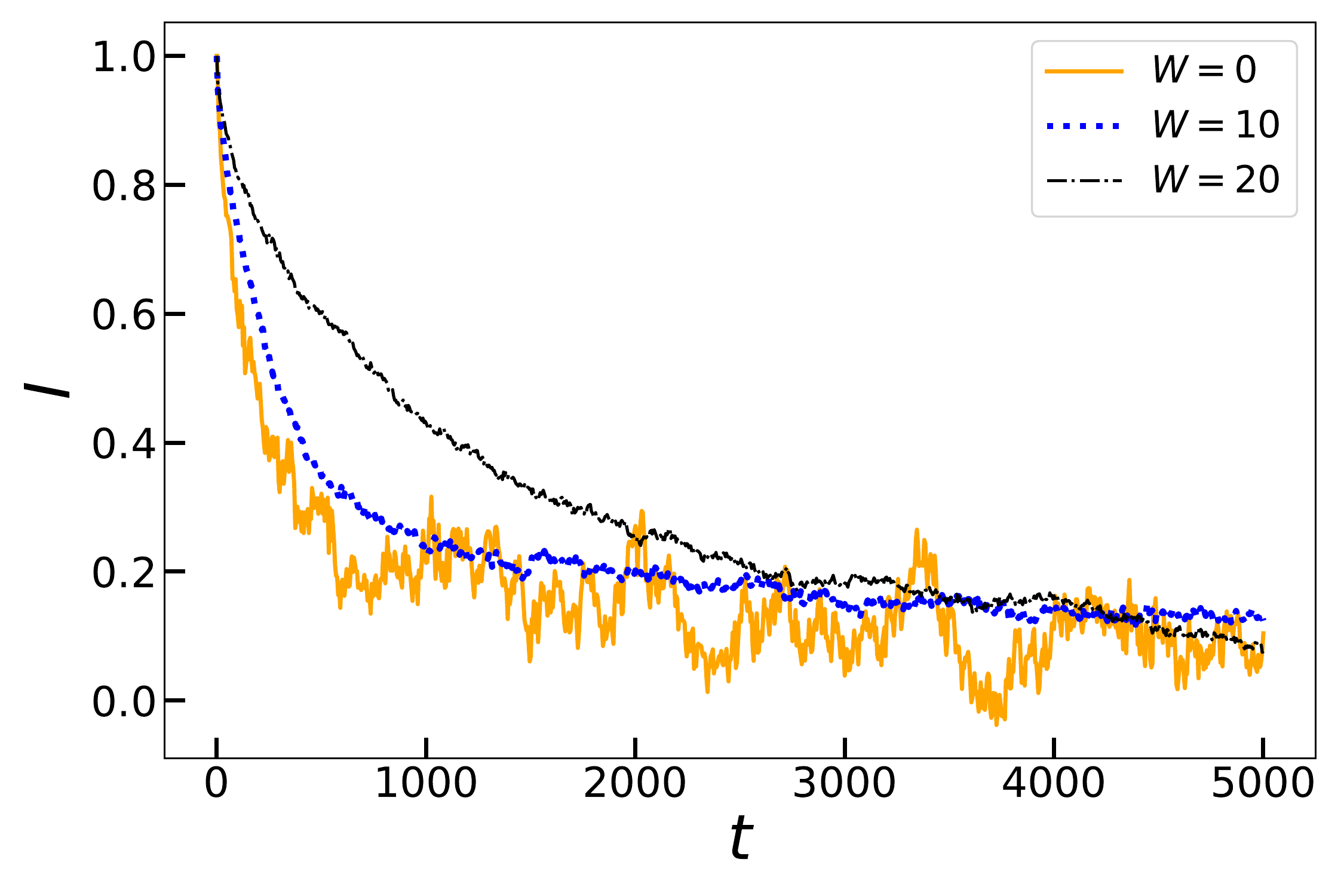}}\\
  \subfigure{\includegraphics[width = 9 cm]{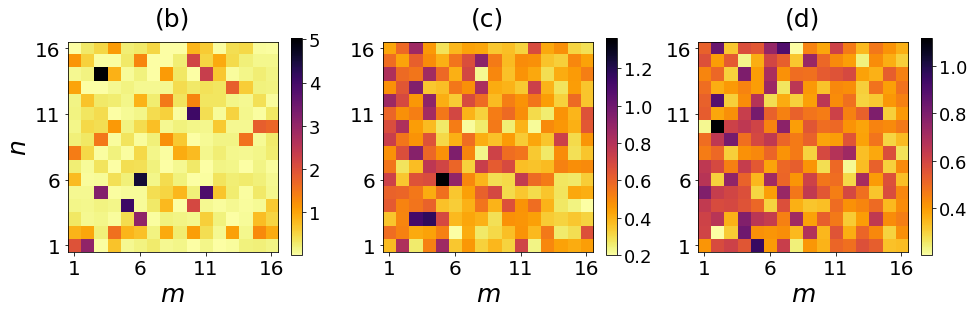}}
    \caption{
Quench dynamics for the $U=10$ path in Fig.~\ref{large_nonlinearity}.
(a) Imbalance $I(t)$ for $W=0,10,20$. (b)-(d) Norm density distribution at final time $t=5000$.
(b) $W=0$, (c) $W=10$, (d) $W=20$. 
All data for $W \neq 0$ are averaged over 20 disorder realizations.
}\label{nu_10_phase_pi}
   \end{figure}
 
For strong nonlinearity $U=10$ the quench path connects the red filled diamond (Gibbs)
and star (non-Gibbs) in Fig.~\ref{large_nonlinearity}.
The evolution outcome is shown in Fig.~\ref{nu_10_phase_pi}.
For $W=0$ we again observe the formation of several (5-6) persistent long-lived strongly localized large amplitude excitations Fig.~\ref{nu_10_phase_pi}{\color{red}(b)}. 
Each of them confines a norm of about 5 so that the imbalance should take values about $0.04...0.1$ which is close to the observed dependence $I(t)$ 
in Fig.~\ref{nu_10_phase_pi}{\color{red}(a)}. 
Increasing the strength of disorder to $W=10$ we still observe remnants of this non-Gibbs dynamics with an additional delay in the relaxation of $I(t)$, while
even stronger disorder $W=20$ reinforces Anderson localization features.

  
\section{Revisiting experimental data} \label{comparison}

\begin{figure}[!htbp]
   \subfigure{\includegraphics[width = 8.5cm]{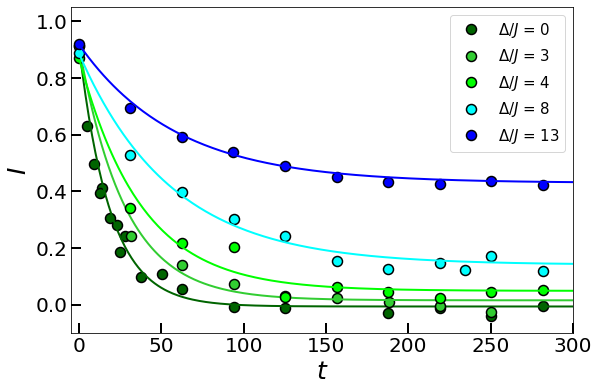}}
   \caption{$I(t)$ for various disorder strength values parameterized through the full width half maximum $\Delta$ (see text for details), as observed in the experiment.
Solid curves guide the eye and correspond to $I(t)=I_0\exp(-t/t_s)+I_{\infty}$. We read $I_{\infty}$ off the last three experimental data points, and $t_s$ from the inset of Fig.~2 in Ref.~\cite{Choi1547}.
}
\label{experiment}
  \end{figure}

The experiments with interacting ultracold bosonic atomic gases loaded into two-dimensional disordered optical potentials
discussed in the introduction result in imbalance curves shown in Fig.~\ref{experiment}. The experimental curves show that the imbalance relaxation slows down with increasing disorder strength, and develops a nonzero asymptotic value.
We note that the disorder potential in the experiment had a Gaussian distribution,
with full width at half maximum  $\Delta$ which corresponds to a variance $\Delta^2/(8\ln(2))$ \cite{Choi1547}. The box disorder which we use in this work has variance $\sigma^2 = W^2/12$, thus we assume 
$\Delta/J=\sqrt{2\ln(2) / 3}W$.
Mapping the experimental setup onto models of interacting bosons results in an interaction strength of $U=24.4$ \cite{Choi1547}. We also note that the experimental records extend to a
largest observation time of $t=300$.

  \begin{figure}[!htbp]
\subfigure[]{\includegraphics[width = 8.2cm]{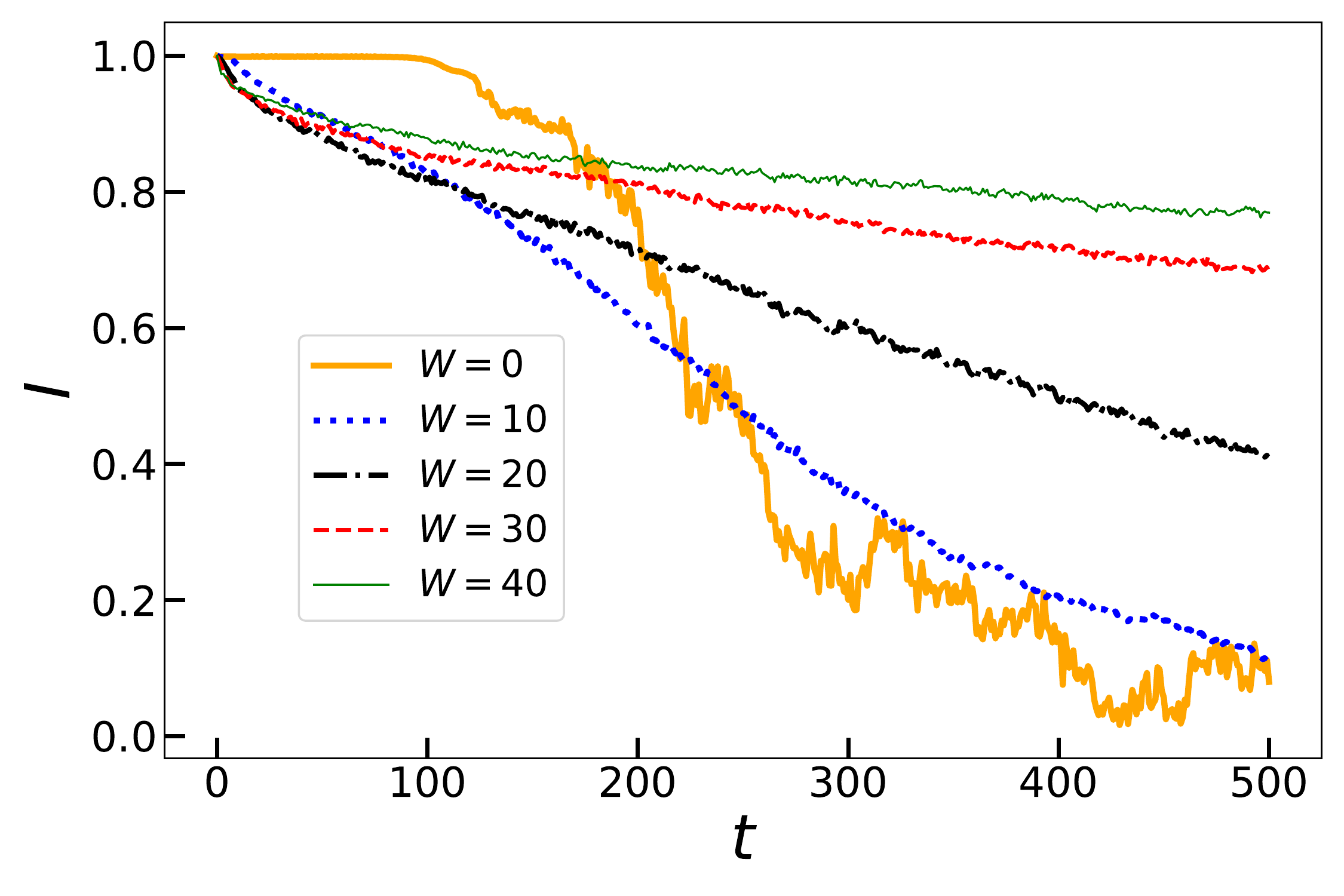}}
\subfigure{\includegraphics[width = 8.5cm]{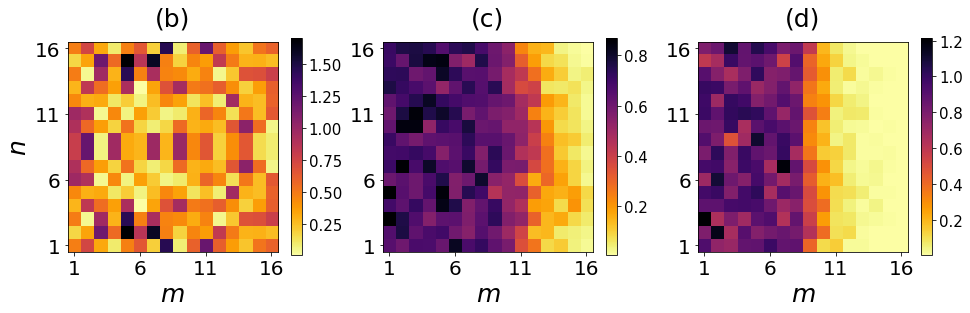}}\\
  \caption{
Gibbs quench dynamics for $U=24.4$. 
(a) Imbalance $I(t)$ for $W=0,10,20,30,40$. (b)-(d) Norm density distribution at final time $t=500$.
(b) $W=0$, (c) $W=20$, (d) $W=40$. 
All data for $W \neq 0$ are averaged over 20 disorder realizations.}\label{comparefig_phase1}    \end{figure}     

In order to compare the experimental results to the DGP dynamics, we use our previous setup with $U=24.4$ and launch the system in the Gibbs regime with initial conditions as
in section \ref{general study}. The Gibbs regime choice follows from the experimental data which show a quick relaxation of the imbalance in the absence of disorder.
Our results are shown in Fig.~\ref{comparefig_phase1}. We observe that the imbalance relaxation is actually {\sl delayed} for the ordered case
compared to the disordered cases. The reason is that the energy shift $U|\psi_{m,n}|^2$ at each excited cite amounts
to $24.4$. Recall that the spectral width of the unexcited lattice part amounts to $\Delta \omega = 8+W$. 
It follows that the excited half of the lattice at $W=0$ is tuned out of resonance (similar to self trapping) with the unexcited one.
At variance, nonzero disorder removes the out-of-resonance feature of the initial state, leading to faster initial decay of the imbalance.
At the same time, stronger disorder hinders full propagation of the excitation into the entire system, which results in a substantial delay of the
imbalance decay at larger time, with almost freezing features at $W=40$. We conclude that the experimental data obtained in the deep quantum
regime show some similarities and differences to the classical runs.


\section{Discussion and Conclusions}\label{concl} 

We investigated the quench expansion dynamics 
of an initially confined state in a two-dimensional Gross-Pitaevskii lattice in the presence of external disorder. 
The expansion dynamics can show qualitatively different outcomes for the imbalance evolution $I(t)$, which depend on the system path
in the control parameter space of the energy and norm densities.
The density space contains a non-Gibbs region. The dynamics in that region leads to strong selftrapping and focusing of potentially large
(compared to the average density) norm on essentially single lattice sites. 
Thermalization in the non-Gibbs regime can or will be substantially
delayed if not completely suppressed, leading to a freezing of the imbalance. On the other side, quenches in the Gibbs regime in general 
result in an imbalance decay, which however can be tremendously postponed by adding strong disorder.

We compared our results to recent experiments with interacting ultracold bosonic atomic gases loaded into two-dimensional disordered optical potentials
 \cite{Choi1547}. Non-Gibbs dynamics is possible for quantum interacting systems as well \cite{PhysRevA.99.023603}. However the experimental
setup reported at most double occupancy per site, which means that the optical potential setup was not capable of trapping more interacting 
atoms per site. Therefore, the impact of non-Gibbs phases can be excluded for the experimental setup.
At the same time we find at least qualitatively similar results for the imbalance relaxation in the Gibbs regime of our system.

\section*{ACKNOWLEDGMENT}
We thank B. L. Altshuler for illuminating discussions while formulating the project.
This work is supported by the Institute for Basic Science, Project Code (IBS-R024-D1). 

\bibliographystyle{apsrev4}
\let\itshape\upshape
\normalem
\bibliography{reference1}

\providecommand{\noopsort}[1]{}\providecommand{\singleletter}[1]{#1}%
\begin{thebibliography}{19}%
\makeatletter
\providecommand \@ifxundefined [1]{%
 \@ifx{#1\undefined}
}%
\providecommand \@ifnum [1]{%
 \ifnum #1\expandafter \@firstoftwo
 \else \expandafter \@secondoftwo
 \fi
}%
\providecommand \@ifx [1]{%
 \ifx #1\expandafter \@firstoftwo
 \else \expandafter \@secondoftwo
 \fi
}%
\providecommand \natexlab [1]{#1}%
\providecommand \enquote  [1]{``#1''}%
\providecommand \bibnamefont  [1]{#1}%
\providecommand \bibfnamefont [1]{#1}%
\providecommand \citenamefont [1]{#1}%
\providecommand \href@noop [0]{\@secondoftwo}%
\providecommand \href [0]{\begingroup \@sanitize@url \@href}%
\providecommand \@href[1]{\@@startlink{#1}\@@href}%
\providecommand \@@href[1]{\endgroup#1\@@endlink}%
\providecommand \@sanitize@url [0]{\catcode `\\12\catcode `\$12\catcode
  `\&12\catcode `\#12\catcode `\^12\catcode `\_12\catcode `\%12\relax}%
\providecommand \@@startlink[1]{}%
\providecommand \@@endlink[0]{}%
\providecommand \url  [0]{\begingroup\@sanitize@url \@url }%
\providecommand \@url [1]{\endgroup\@href {#1}{\urlprefix }}%
\providecommand \urlprefix  [0]{URL }%
\providecommand \Eprint [0]{\href }%
\providecommand \doibase [0]{http://dx.doi.org/}%
\providecommand \selectlanguage [0]{\@gobble}%
\providecommand \bibinfo  [0]{\@secondoftwo}%
\providecommand \bibfield  [0]{\@secondoftwo}%
\providecommand \translation [1]{[#1]}%
\providecommand \BibitemOpen [0]{}%
\providecommand \bibitemStop [0]{}%
\providecommand \bibitemNoStop [0]{.\EOS\space}%
\providecommand \EOS [0]{\spacefactor3000\relax}%
\providecommand \BibitemShut  [1]{\csname bibitem#1\endcsname}%
\let\auto@bib@innerbib\@empty
\bibitem [{\citenamefont {Polkovnikov}\ \emph {et~al.}(2011)\citenamefont
  {Polkovnikov}, \citenamefont {Sengupta}, \citenamefont {Silva},\ and\
  \citenamefont {Vengalattore}}]{Polkovnikov_2011}%
  \BibitemOpen
  \bibfield  {author} {\bibinfo {author} {\bibfnamefont {A.}~\bibnamefont
  {Polkovnikov}}, \bibinfo {author} {\bibfnamefont {K.}~\bibnamefont
  {Sengupta}}, \bibinfo {author} {\bibfnamefont {A.}~\bibnamefont {Silva}}, \
  and\ \bibinfo {author} {\bibfnamefont {M.}~\bibnamefont {Vengalattore}},\
  }\bibfield  {title} {\enquote {\bibinfo {title} {Colloquium: Nonequilibrium
  dynamics of closed interacting quantum systems},}\ }\href {\doibase
  10.1103/RevModPhys.83.863} {\bibfield  {journal} {\bibinfo  {journal} {Rev.
  Mod. Phys.}\ }\textbf {\bibinfo {volume} {83}},\ \bibinfo {pages} {863}
  (\bibinfo {year} {2011})}\BibitemShut {NoStop}%
\bibitem [{\citenamefont {Choi}\ \emph {et~al.}(2016)\citenamefont {Choi},
  \citenamefont {Hild}, \citenamefont {Zeiher}, \citenamefont {Schau{\ss}},
  \citenamefont {Rubio-Abadal}, \citenamefont {Yefsah}, \citenamefont
  {Khemani}, \citenamefont {Huse}, \citenamefont {Bloch},\ and\ \citenamefont
  {Gross}}]{Choi1547}%
  \BibitemOpen
  \bibfield  {author} {\bibinfo {author} {\bibfnamefont {J.-y.}\ \bibnamefont
  {Choi}}, \bibinfo {author} {\bibfnamefont {S.}~\bibnamefont {Hild}}, \bibinfo
  {author} {\bibfnamefont {J.}~\bibnamefont {Zeiher}}, \bibinfo {author}
  {\bibfnamefont {P.}~\bibnamefont {Schau{\ss}}}, \bibinfo {author}
  {\bibfnamefont {A.}~\bibnamefont {Rubio-Abadal}}, \bibinfo {author}
  {\bibfnamefont {T.}~\bibnamefont {Yefsah}}, \bibinfo {author} {\bibfnamefont
  {V.}~\bibnamefont {Khemani}}, \bibinfo {author} {\bibfnamefont {D.~A.}\
  \bibnamefont {Huse}}, \bibinfo {author} {\bibfnamefont {I.}~\bibnamefont
  {Bloch}}, \ and\ \bibinfo {author} {\bibfnamefont {C.}~\bibnamefont
  {Gross}},\ }\bibfield  {title} {\enquote {\bibinfo {title} {Exploring the
  many-body localization transition in two dimensions},}\ }\href {\doibase
  10.1126/science.aaf8834} {\bibfield  {journal} {\bibinfo  {journal}
  {Science}\ }\textbf {\bibinfo {volume} {352}},\ \bibinfo {pages} {1547}
  (\bibinfo {year} {2016})}\BibitemShut {NoStop}%
\bibitem [{\citenamefont {Yan}\ \emph {et~al.}(2017)\citenamefont {Yan},
  \citenamefont {Hui}, \citenamefont {Rigol},\ and\ \citenamefont
  {Scarola}}]{Yan_2017}%
  \BibitemOpen
  \bibfield  {author} {\bibinfo {author} {\bibfnamefont {M.}~\bibnamefont
  {Yan}}, \bibinfo {author} {\bibfnamefont {H.-Y.}\ \bibnamefont {Hui}},
  \bibinfo {author} {\bibfnamefont {M.}~\bibnamefont {Rigol}}, \ and\ \bibinfo
  {author} {\bibfnamefont {V.~W.}\ \bibnamefont {Scarola}},\ }\bibfield
  {title} {\enquote {\bibinfo {title} {Equilibration dynamics of strongly
  interacting bosons in 2d lattices with disorder},}\ }\href {\doibase
  10.1103/PhysRevLett.119.073002} {\bibfield  {journal} {\bibinfo  {journal}
  {Phys. Rev. Lett.}\ }\textbf {\bibinfo {volume} {119}},\ \bibinfo {pages}
  {073002} (\bibinfo {year} {2017})}\BibitemShut {NoStop}%
\bibitem [{\citenamefont {Urbanek}\ and\ \citenamefont
  {Sold{\'a}n}(2018)}]{urbanek2018}%
  \BibitemOpen
  \bibfield  {author} {\bibinfo {author} {\bibfnamefont {M.}~\bibnamefont
  {Urbanek}}\ and\ \bibinfo {author} {\bibfnamefont {P.}~\bibnamefont
  {Sold{\'a}n}},\ }\bibfield  {title} {\enquote {\bibinfo {title}
  {Equilibration in two-dimensional bose systems with disorders},}\ }\href@noop
  {} {\bibfield  {journal} {\bibinfo  {journal} {The European Physical Journal
  D}\ }\textbf {\bibinfo {volume} {72}},\ \bibinfo {pages} {114} (\bibinfo
  {year} {2018})}\BibitemShut {NoStop}%
\bibitem [{\citenamefont {Dutta}\ \emph {et~al.}(2015)\citenamefont {Dutta},
  \citenamefont {Gajda}, \citenamefont {Hauke}, \citenamefont {Lewenstein},
  \citenamefont {L{\"u}hmann}, \citenamefont {Malomed}, \citenamefont
  {Sowi{\'{n}}ski},\ and\ \citenamefont {Zakrzewski}}]{Dutta_2015}%
  \BibitemOpen
  \bibfield  {author} {\bibinfo {author} {\bibfnamefont {O.}~\bibnamefont
  {Dutta}}, \bibinfo {author} {\bibfnamefont {M.}~\bibnamefont {Gajda}},
  \bibinfo {author} {\bibfnamefont {P.}~\bibnamefont {Hauke}}, \bibinfo
  {author} {\bibfnamefont {M.}~\bibnamefont {Lewenstein}}, \bibinfo {author}
  {\bibfnamefont {D.-S.}\ \bibnamefont {L{\"u}hmann}}, \bibinfo {author}
  {\bibfnamefont {B.~A.}\ \bibnamefont {Malomed}}, \bibinfo {author}
  {\bibfnamefont {T.}~\bibnamefont {Sowi{\'{n}}ski}}, \ and\ \bibinfo {author}
  {\bibfnamefont {J.}~\bibnamefont {Zakrzewski}},\ }\bibfield  {title}
  {\enquote {\bibinfo {title} {Non-standard hubbard models in optical lattices:
  a review},}\ }\href {\doibase 10.1088/0034-4885/78/6/066001} {\bibfield
  {journal} {\bibinfo  {journal} {Reports on Progress in Physics}\ }\textbf
  {\bibinfo {volume} {78}},\ \bibinfo {pages} {066001} (\bibinfo {year}
  {2015})}\BibitemShut {NoStop}%
\bibitem [{\citenamefont {Kevrekidis}(2009)}]{kevrekidis2009discrete}%
  \BibitemOpen
  \bibfield  {author} {\bibinfo {author} {\bibfnamefont {P.~G.}\ \bibnamefont
  {Kevrekidis}},\ }\href@noop {} {\emph {\bibinfo {title} {The discrete
  nonlinear Schr{\"o}dinger equation: mathematical analysis, numerical
  computations and physical perspectives}}},\ Vol.\ \bibinfo {volume} {232}\
  (\bibinfo  {publisher} {Springer Science \& Business Media},\ \bibinfo {year}
  {2009})\BibitemShut {NoStop}%
\bibitem [{\citenamefont {Abanin}\ \emph {et~al.}(2019)\citenamefont {Abanin},
  \citenamefont {Altman}, \citenamefont {Bloch},\ and\ \citenamefont
  {Serbyn}}]{RevModPhys.91.021001}%
  \BibitemOpen
  \bibfield  {author} {\bibinfo {author} {\bibfnamefont {D.~A.}\ \bibnamefont
  {Abanin}}, \bibinfo {author} {\bibfnamefont {E.}~\bibnamefont {Altman}},
  \bibinfo {author} {\bibfnamefont {I.}~\bibnamefont {Bloch}}, \ and\ \bibinfo
  {author} {\bibfnamefont {M.}~\bibnamefont {Serbyn}},\ }\bibfield  {title}
  {\enquote {\bibinfo {title} {Colloquium: Many-body localization,
  thermalization, and entanglement},}\ }\href {\doibase
  10.1103/RevModPhys.91.021001} {\bibfield  {journal} {\bibinfo  {journal}
  {Rev. Mod. Phys.}\ }\textbf {\bibinfo {volume} {91}},\ \bibinfo {pages}
  {021001} (\bibinfo {year} {2019})}\BibitemShut {NoStop}%
\bibitem [{\citenamefont {Mithun}\ \emph {et~al.}(2018)\citenamefont {Mithun},
  \citenamefont {Kati}, \citenamefont {Danieli},\ and\ \citenamefont
  {Flach}}]{Mithun_2018}%
  \BibitemOpen
  \bibfield  {author} {\bibinfo {author} {\bibfnamefont {T.}~\bibnamefont
  {Mithun}}, \bibinfo {author} {\bibfnamefont {Y.}~\bibnamefont {Kati}},
  \bibinfo {author} {\bibfnamefont {C.}~\bibnamefont {Danieli}}, \ and\
  \bibinfo {author} {\bibfnamefont {S.}~\bibnamefont {Flach}},\ }\bibfield
  {title} {\enquote {\bibinfo {title} {Weakly nonergodic dynamics in the
  gross-pitaevskii lattice},}\ }\href {\doibase 10.1103/PhysRevLett.120.184101}
  {\bibfield  {journal} {\bibinfo  {journal} {Phys. Rev. Lett.}\ }\textbf
  {\bibinfo {volume} {120}},\ \bibinfo {pages} {184101} (\bibinfo {year}
  {2018})}\BibitemShut {NoStop}%
\bibitem [{\citenamefont {Cherny}\ \emph {et~al.}(2019)\citenamefont {Cherny},
  \citenamefont {Engl},\ and\ \citenamefont {Flach}}]{PhysRevA.99.023603}%
  \BibitemOpen
  \bibfield  {author} {\bibinfo {author} {\bibfnamefont {A.~Y.}\ \bibnamefont
  {Cherny}}, \bibinfo {author} {\bibfnamefont {T.}~\bibnamefont {Engl}}, \ and\
  \bibinfo {author} {\bibfnamefont {S.}~\bibnamefont {Flach}},\ }\bibfield
  {title} {\enquote {\bibinfo {title} {Non-gibbs states on a bose-hubbard
  lattice},}\ }\href {\doibase 10.1103/PhysRevA.99.023603} {\bibfield
  {journal} {\bibinfo  {journal} {Phys. Rev. A}\ }\textbf {\bibinfo {volume}
  {99}},\ \bibinfo {pages} {023603} (\bibinfo {year} {2019})}\BibitemShut
  {NoStop}%
\bibitem [{\citenamefont {Rasmussen}\ \emph {et~al.}(2000)\citenamefont
  {Rasmussen}, \citenamefont {Cretegny}, \citenamefont {Kevrekidis},\ and\
  \citenamefont {Gr\o{}nbech-Jensen}}]{Rasmussen_2000}%
  \BibitemOpen
  \bibfield  {author} {\bibinfo {author} {\bibfnamefont {K.~O.}\ \bibnamefont
  {Rasmussen}}, \bibinfo {author} {\bibfnamefont {T.}~\bibnamefont {Cretegny}},
  \bibinfo {author} {\bibfnamefont {P.~G.}\ \bibnamefont {Kevrekidis}}, \ and\
  \bibinfo {author} {\bibfnamefont {N.}~\bibnamefont {Gr\o{}nbech-Jensen}},\
  }\bibfield  {title} {\enquote {\bibinfo {title} {Statistical mechanics of a
  discrete nonlinear system},}\ }\href {\doibase 10.1103/PhysRevLett.84.3740}
  {\bibfield  {journal} {\bibinfo  {journal} {Phys. Rev. Lett.}\ }\textbf
  {\bibinfo {volume} {84}},\ \bibinfo {pages} {3740} (\bibinfo {year}
  {2000})}\BibitemShut {NoStop}%
\bibitem [{\citenamefont {Johansson}\ and\ \citenamefont
  {Rasmussen}(2004)}]{Johansson_2004}%
  \BibitemOpen
  \bibfield  {author} {\bibinfo {author} {\bibfnamefont {M.}~\bibnamefont
  {Johansson}}\ and\ \bibinfo {author} {\bibfnamefont {K.~O.}\ \bibnamefont
  {Rasmussen}},\ }\bibfield  {title} {\enquote {\bibinfo {title} {Statistical
  mechanics of general discrete nonlinear schr\"odinger models: Localization
  transition and its relevance for klein-gordon lattices},}\ }\href {\doibase
  10.1103/PhysRevE.70.066610} {\bibfield  {journal} {\bibinfo  {journal} {Phys.
  Rev. E}\ }\textbf {\bibinfo {volume} {70}},\ \bibinfo {pages} {066610}
  (\bibinfo {year} {2004})}\BibitemShut {NoStop}%
\bibitem [{\citenamefont {Yoshida}(1990)}]{Yoshida:1990aa}%
  \BibitemOpen
  \bibfield  {author} {\bibinfo {author} {\bibfnamefont {H.}~\bibnamefont
  {Yoshida}},\ }\bibfield  {title} {\enquote {\bibinfo {title} {Construction of
  higher order symplectic integrators},}\ }\href@noop {} {\bibfield  {journal}
  {\bibinfo  {journal} {Physics letters A}\ }\textbf {\bibinfo {volume}
  {150}},\ \bibinfo {pages} {262} (\bibinfo {year} {1990})}\BibitemShut
  {NoStop}%
\bibitem [{\citenamefont {McLachlan}(1995)}]{McLachlan:1995aa}%
  \BibitemOpen
  \bibfield  {author} {\bibinfo {author} {\bibfnamefont {R.~I.}\ \bibnamefont
  {McLachlan}},\ }\bibfield  {title} {\enquote {\bibinfo {title} {Composition
  methods in the presence of small parameters},}\ }\href@noop {} {\bibfield
  {journal} {\bibinfo  {journal} {BIT Numerical Mathematics}\ }\textbf
  {\bibinfo {volume} {35}},\ \bibinfo {pages} {258} (\bibinfo {year}
  {1995})}\BibitemShut {NoStop}%
\bibitem [{\citenamefont {Laskar}\ and\ \citenamefont
  {Robutel}(2001)}]{laskar_2001}%
  \BibitemOpen
  \bibfield  {author} {\bibinfo {author} {\bibfnamefont {J.}~\bibnamefont
  {Laskar}}\ and\ \bibinfo {author} {\bibfnamefont {P.}~\bibnamefont
  {Robutel}},\ }\bibfield  {title} {\enquote {\bibinfo {title} {High order
  symplectic integrators for perturbed hamiltonian systems},}\ }\href
  {https://doi.org/10.1023/A:1012098603882} {\bibfield  {journal} {\bibinfo
  {journal} {Celestial Mechanics and Dynamical Astronomy}\ }\textbf {\bibinfo
  {volume} {80}},\ \bibinfo {pages} {39} (\bibinfo {year} {2001})}\BibitemShut
  {NoStop}%
\bibitem [{\citenamefont {Skokos}\ \emph {et~al.}(2009)\citenamefont {Skokos},
  \citenamefont {Krimer}, \citenamefont {Komineas},\ and\ \citenamefont
  {Flach}}]{Skokos:2009aa}%
  \BibitemOpen
  \bibfield  {author} {\bibinfo {author} {\bibfnamefont {C.}~\bibnamefont
  {Skokos}}, \bibinfo {author} {\bibfnamefont {D.}~\bibnamefont {Krimer}},
  \bibinfo {author} {\bibfnamefont {S.}~\bibnamefont {Komineas}}, \ and\
  \bibinfo {author} {\bibfnamefont {S.}~\bibnamefont {Flach}},\ }\bibfield
  {title} {\enquote {\bibinfo {title} {Delocalization of wave packets in
  disordered nonlinear chains},}\ }\href@noop {} {\bibfield  {journal}
  {\bibinfo  {journal} {Physical Review E}\ }\textbf {\bibinfo {volume} {79}},\
  \bibinfo {pages} {056211} (\bibinfo {year} {2009})}\BibitemShut {NoStop}%
\bibitem [{\citenamefont {Skokos}\ \emph {et~al.}(2014)\citenamefont {Skokos},
  \citenamefont {Krimer}, \citenamefont {Komineas},\ and\ \citenamefont
  {Flach}}]{Skokos:2014aa}%
  \BibitemOpen
  \bibfield  {author} {\bibinfo {author} {\bibfnamefont {C.}~\bibnamefont
  {Skokos}}, \bibinfo {author} {\bibfnamefont {D.}~\bibnamefont {Krimer}},
  \bibinfo {author} {\bibfnamefont {S.}~\bibnamefont {Komineas}}, \ and\
  \bibinfo {author} {\bibfnamefont {S.}~\bibnamefont {Flach}},\ }\bibfield
  {title} {\enquote {\bibinfo {title} {Erratum: Delocalization of wave packets
  in disordered nonlinear chains {$[$}phys. rev. e 79, 056211 (2009){$]$}},}\
  }\href@noop {} {\bibfield  {journal} {\bibinfo  {journal} {Physical Review
  E}\ }\textbf {\bibinfo {volume} {89}},\ \bibinfo {pages} {029907} (\bibinfo
  {year} {2014})}\BibitemShut {NoStop}%
\bibitem [{\citenamefont {Danieli}\ \emph {et~al.}(2019)\citenamefont
  {Danieli}, \citenamefont {Manda}, \citenamefont {Thudiyangal},\ and\
  \citenamefont {Skokos}}]{Carlo_2019}%
  \BibitemOpen
  \bibfield  {author} {\bibinfo {author} {\bibfnamefont {C.}~\bibnamefont
  {Danieli}}, \bibinfo {author} {\bibfnamefont {B.~M.}\ \bibnamefont {Manda}},
  \bibinfo {author} {\bibfnamefont {M.}~\bibnamefont {Thudiyangal}}, \ and\
  \bibinfo {author} {\bibfnamefont {C.}~\bibnamefont {Skokos}},\ }\bibfield
  {title} {\enquote {\bibinfo {title} {Computational efficiency of numerical
  integration methods for the tangent dynamics of many-body hamiltonian systems
  in one and two spatial dimensions},}\ }\href {\doibase
  http://dx.doi.org/10.3934/mine.2019.3.447} {\bibfield  {journal} {\bibinfo
  {journal} {Mathematics in Engineering}\ }\textbf {\bibinfo {volume} {1}},\
  \bibinfo {pages} {447} (\bibinfo {year} {2019})}\BibitemShut {NoStop}%
\bibitem [{\citenamefont {Anderson}(1958)}]{Anderson_1958}%
  \BibitemOpen
  \bibfield  {author} {\bibinfo {author} {\bibfnamefont {P.~W.}\ \bibnamefont
  {Anderson}},\ }\bibfield  {title} {\enquote {\bibinfo {title} {Absence of
  diffusion in certain random lattices},}\ }\href {\doibase
  10.1103/PhysRev.109.1492} {\bibfield  {journal} {\bibinfo  {journal} {Phys.
  Rev.}\ }\textbf {\bibinfo {volume} {109}},\ \bibinfo {pages} {1492} (\bibinfo
  {year} {1958})}\BibitemShut {NoStop}%
\bibitem [{\citenamefont {Rumpf}(2004)}]{Rumpf_2004}%
  \BibitemOpen
  \bibfield  {author} {\bibinfo {author} {\bibfnamefont {B.}~\bibnamefont
  {Rumpf}},\ }\bibfield  {title} {\enquote {\bibinfo {title} {Simple
  statistical explanation for the localization of energy in nonlinear lattices
  with two conserved quantities},}\ }\href {\doibase
  10.1103/PhysRevE.69.016618} {\bibfield  {journal} {\bibinfo  {journal} {Phys.
  Rev. E}\ }\textbf {\bibinfo {volume} {69}},\ \bibinfo {pages} {016618}
  (\bibinfo {year} {2004})}\BibitemShut {NoStop}%
\end{thebibliography}%

\end{document}